\documentclass[10pt, showpacs,preprintnumbers, pra]{revtex4}
\usepackage{graphicx}
\usepackage{floatrow}
\usepackage[caption=false]{subfig}
\usepackage{xcolor}
\usepackage{amsmath}
\usepackage{amssymb}
\usepackage{mleftright}
\usepackage{dsfont}
\usepackage{floatrow}
\usepackage{mathtools}
\mleftright
\def\bc{\begin{center}}
\def\ec{\end{center}}

\def\beq{\begin{equation}}
\def\eeq{\end{equation}}

\begin{document}

\title{Dynamics of a qubit-oscillator system with periodically varying coupling}
\author{Mirko Amico$^{1,2}$ and Roman Ya. Kezerashvili$^{1,2}$}
\affiliation{\mbox{$^{1}$Physics Department, New York City College
of Technology, The City University of New York,} \\
Brooklyn, NY 11201, USA \\
\mbox{$^{2}$The Graduate School and University Center, The
City University of New York,} \\
New York, NY 10016, USA}

\begin{abstract}
The dynamics of qubits coupled to a harmonic oscillator with time-periodic coupling is investigated in the framework of Floquet theory. This system can be used to model nonadiabatic phenomena that require a periodic modulation of the qubit/oscillator coupling. The case of a single qubit coupled to a resonator populated with $n= 0,1$ photons is explicitly treated. The time-dependent Schr\"{o}dinger equation describing the system's dynamics is solved within the Floquet formalism and a perturbative approach in the time- and Laplace-domain. Good quantitative agreement is found between the analytical and numerical calculations within the Floquet approach, making it the most promising candidate for the study of time-periodic problems. Nonetheless, the time- or Laplace-domain perturbative approaches can be used in the presence of aperiodic time-dependent terms in the Hamiltonian.

\end{abstract}


\maketitle

\section{Introduction}
\label{intro}

The dynamics of quantum systems, which are periodically driven by low- and high-frequency field, has been widely investigated in the framework of the Floquet approach \cite{rahav, shtoff, goldman, holthaus, chu}. The Floquet formalism was introduced in Ref. \cite{floquet} to simplify the solution of ordinary differential equations with terms that show a certain periodicity in time. More generally, it allows to consider the case of solutions of a linear partial differential equation periodic with respect to several variables, for example, periodic with respect to a crystal lattice or time \cite{eastham, daleckii, ashcroft}. Clearly, this method can be applied to the Schr\"{o}dinger equation for the study of the time-evolution of quantum system with periodic Hamiltonian \cite{shirley, sambe, levante, leskes, grifoni}. The approach turns the problem of solving a differential equation into the problem of finding the eigenvalues and the eigenvectors of a matrix. Depending on the problem at hand, this can make it easier to find its solution. 

Here, we consider a system composed of $N$ two-level systems, also called qubits, coupled to a harmonic oscillator. This is a good model for many physical systems such as atoms coupled to the electromagnetic field inside a cavity or superconducting circuits with non-linear elements coupled to a superconducting coplanar waveguide. More specifically, we consider the case of a superconducting circuit system where the qubit/resonator coupling is modulated periodically. As found in Ref. \cite{shapiro, zhukov, amico1, remizov, amico2, amico3, amico4}, if the coupling is nonadiabatically modulated, the qubits and photons in the resonator can be excited from the ground state. Furthermore, periodic modulation of the coupling can greatly increase the probability of excitation of both qubits and resonator. This phenomenon, which involves the creation of excitation from the quantum vacuum and is caused by the nonadiabatic change of the boundary conditions of the system, is known as dynamical Lamb effect (DLE) \cite{dle}. 

Since the periodic modulation of the coupling is required to enhance the effects of the DLE, one can consider the Floquet approach to study the time-evolution of the system. Refs. \cite{son, deng, pirkka} present applications of the Floquet formalism to the study of the dynamics of superconducting circuits. While in Refs. \cite{marinescu, gavrila} a perturbation theory for the Floquet states and the quasienergies in terms of the states of the time-averaged problem is developed. As it turns out, the Floquet approach provides a framework where analytical calculations can be performed and, despite the approximations, give results which are in good agreement with the numerical calculations. The effects of the DLE is studied by considering the time-evolution of the state $ \lvert e, 1 \rangle $. In fact, as discussed in \cite{amico4}, the latter can only be reached from the ground state if the counter-rotating terms $\hat{\sigma}^+ \hat{a}$ and $\hat{\sigma}^- \hat{a}^{\dagger}$ in the interaction Hamiltonian of the system become relevant. This can happen, for example, when the qubit/resonator coupling is modulated at the sum frequency of the qubit and the resonator transition frequencies. 

In our previous work \cite{amico2, amico3}, we have investigated the dynamics of a system of $N$ qubits coupled to a resonator using different approaches. In Ref. \cite{amico2}, the time-dependent Schr\"{o}dinger equation is solved directly in the time-domain using a perturbative approach and considering an averaged time-independent coupling. In Ref. \cite{amico3} the Laplace transform is used to show how ordinary differential equations with a complicated time-dependence can easily be solved in the Laplace-domain. We briefly introduce these different approaches and compare the results obtained using these methods to the Floquet approach.

In this work, we consider the case of a single qubit coupled to a resonator. The time evolution of the state $ \lvert e, 1 \rangle $ can then easily be calculated both analytically and numerically to show the effects of the DLE. The different methods described above are used to study the dynamics of the system and their results are compared. Overall, all methods give comparable results and the solution obtained with each of them shows the same qualitative features. Good quantitative agreement is found between the analytical and numerical calculations within the Floquet approach, making it the most promising candidate for the study of time-periodic problems.

The article is organized in the following way. Sec. \ref{sup} defines the Hamiltonian of the system considered. The general case of $N$ qubits coupled to a resonator is described. The Floquet formalism is introduced and applied to the case of a single qubit coupled to a resonator in Sec. \ref{floquet}. Analytical and numerical calculations within the Floquet approach are then compared. In Sec. \ref{comp}, other analytical and numerical methods are presented. These include the perturbative integration of the Schr\"{o}dinger equation in time- and Laplace-domain. The time-evolution of the $\lvert e, 1 \rangle$ state is explicitly calculated in the framework of these methods. A comparison of all the results obtained within the different approaches is given in Sec. \ref{res}. Conclusions follow in Sec. \ref{conc}.

\section{$N$ superconducting qubits coupled to a resonator}
\label{sup}

Let us consider a system of $N$ superconducting qubits coupled to a resonator with a time-dependent coupling. This system is well described by the Hamiltonian of $N$ two-level systems coupled to a single-mode of the electromagnetic field with a time-dependent coupling, which is used to describe an atom interacting with the electromagnetic field of a cavity with variable atom/cavity coupling strength

\begin{equation}
\hat{H} (t)= \hbar \omega _{r}\hat{a}^{\dagger }\hat{a} + \sum_i^N \left[ \hbar \omega _{0}^i \hat{\sigma}^{+}_i \hat{\sigma}^{-}_i +\hbar g(t) \hat{\sigma}_{x}^i \left( \hat{a}^{\dagger }+\hat{a}\right) \right] . 
\label{H_qub_em}
\end{equation}

\noindent
For the case of a superconducting circuit setup, $\omega _{0}^i$ is the transition frequency of the $i$-th qubit, $\omega _{r}$ is the frequency of the photons in the resonator, $\hat{\sigma}^{+}_i = \frac{\hat{\sigma}_x^i + i \hat{\sigma}_y^i }{2}$, $\hat{\sigma}^{-}_i = \frac{\hat{\sigma}_x^i - i \hat{\sigma}_y^i }{2}$ and $\hat{a}^{\dagger }$, $\hat{a}$ are the creation and
annihilation operators for excitations of qubits and photons, respectively, $\hat{\sigma}_{x}^i$, $\hat{\sigma}_{y}^i$ and $\hat{\sigma}_{z}^i$ are the Pauli $x$, $y$ and $z$ operators for each qubit, while $g(t)$ is the time-dependent coupling strength between the qubit and the resonator. Differently from the case of atoms in a cavity, the parameters of superconducting circuits can be engineered over a wide range of values. This allows, for instance, to increase or decrease the coupling strength between qubits and the resonator to very low or high values, effectively turning "off" or "on" their coupling. Let us mention that a possible experimental setup where this can be achieved was proposed in Ref. \cite{amico4}.

%
%
%

The Hamiltonian (\ref{H_qub_em}) can be split into two parts:

\begin{equation}
\hat{H}(t)=\hat{H}_{0}+\hat{H}_{I}(t),  
\label{H_split}
\end{equation}
where 

\begin{eqnarray}
\hat{H}_{0} &=&\hbar \omega _{r}\hat{a}^{\dagger }\hat{a}+\sum_{i}^{N}\hbar
\omega _{0}^{i}\hat{\sigma}_{i}^{+}\hat{\sigma}_{i}^{-} \\
\hat{H}_{I}(t) &=&\sum_{i}^{N}\left[ \hbar g(t)\hat{\sigma}_{x}^{i}\left( 
\hat{a}^{\dagger }+\hat{a}\right) \right] 
\end{eqnarray}

\noindent are a time-independent and a time-dependent term, respectively. As
discussed and demonstrated in Refs. \cite{shapiro, amico2}, the DLE is
maximum when the qubit/resonator coupling is modulated periodically.
Furthermore, in Ref. \cite{amico4} we found that a sinusoidal modulation of
the right frequency can be used. Therefore, we take the qubit/resonator
coupling as

\begin{equation}
\begin{split}
\label{g_sin}
g(t) = g_0  \cos \left( \varpi_s t   \right) ,
\end{split}
\end{equation}

\noindent
where $g_0$ is the qubit/resonator coupling strength and ${\varpi_s}$ is the frequency of switching of the coupling. Thus, the time-dependent Hamiltonian term is also periodic $\hat{H}_{I}(t) = \hat{H}_{I}(t+T)$, with period $T =  \frac{2\pi}{\varpi_s}$.

\section{Floquet theory}
\label{floquet}

Given a quantum system described by a Hamiltonian periodic in time with a period $T$, that is $\hat{H} (t) = \hat{H} (t+T) $, one can investigate the dynamics of the system within the Floquet approach \cite{shirley, grifoni}. Let us now consider the Schr\"{o}dinger equation with the Hamiltonian (\ref{H_split}), which describes the time-evolution of the state $\lvert \psi (t) \rangle$ of a system

\begin{equation}
i \hbar \frac{\partial}{\partial t} \lvert \psi (t) \rangle = \hat{H} (t)  \lvert \psi (t) \rangle .
\label{schro}
\end{equation}

\noindent
Since $\hat{H} (t)$ is periodic, Eq. (\ref{schro}) is a differential equation with periodic coefficients. Thus, one can find its solutions using the Floquet theorem \cite{floquet} in the following form

\begin{equation}
\lvert \psi_{\alpha} (t) \rangle =  \lvert \phi_{\alpha} (t) \rangle e^{-i  \frac{\epsilon_{\alpha} t}{\hbar}} ,
\label{floq}
\end{equation}
 
\noindent 
where $ \lvert \phi_{\alpha} (t) \rangle$ is called \textit{Floquet mode} and is a periodic function with period $T$ and $ \epsilon_{\alpha}$ is called \textit{quasienergy} or \textit{Floquet characteristic exponent}. Clearly, different values of the quasienergy $\epsilon_{\alpha '} = \epsilon_{\alpha} + n \frac{ 2 \pi \hbar}{T}$, where $n=0, \pm 1, \pm 2, ... $, correspond to the same solution $\lvert \psi_{\alpha} (t) \rangle$.


It is important to note that the Floquet modes and the quasienergies are the eigenfunctions and eigenvalues, respectively, of the operator $\hat{\mathcal{H}}(t) = \hat{H}(t) - i \hbar \frac{\partial}{\partial t}$

\begin{equation}
\hat{\mathcal{H}}(t) \lvert \phi_{\alpha} (t) \rangle = \epsilon_{\alpha}  \lvert \phi_{\alpha} (t) \rangle .
\label{eig}
\end{equation}

\noindent
This can be seen by substituting the general solution given in Eq. (\ref{floq}) into the Schr\"{o}dinger equation (\ref{schro}). Then, Eq. (\ref{eig}) provides an alternative way of determining the state $\lvert \psi (t) \rangle$ of the system. To demonstrate this, let us consider the decomposition of the generic state $\lvert \psi (t) \rangle$ in terms of Floquet modes

\begin{equation}
\lvert \psi (t) \rangle  = \sum_{\alpha} c_\alpha \lvert \phi_{\alpha} (t) \rangle e^{\frac{-i \epsilon_{\alpha} t}{\hbar}},
\label{deco}
\end{equation}

\noindent
where the coefficients $c_\alpha = \langle \psi (0) \lvert \phi_\alpha (0) \rangle$ quantify the overlap of the wavefunction with the Floquet modes at time $t=0$. Following Refs. \cite{creffield, qutip1}, one can find the Floquet modes by noting that they are eigenfunctions of the time-evolution operator $U \left(t, t_0 \right)$. In fact, the Schr\"{o}dinger equation written in terms of the time-evolution operator,

\begin{equation}
U \left(t_0 + T, t_0 \right) \lvert \psi (t_0) \rangle = \lvert \psi (t_0 + T) \rangle ,
\label{time_evo}
\end{equation}

\noindent
can be rewritten in terms of a generic Floquet mode as 

\begin{equation}
U \left(t_0 + T, t_0 \right)  \lvert \phi_{\alpha} (t_0) \rangle e^{\frac{-i \epsilon_{\alpha} t_0}{\hbar}} =  \lvert \phi_{\alpha} (t_0 + T) \rangle e^{\frac{-i \epsilon_{\alpha}}{\hbar}  \left( t_0 + T \right) } .
\label{time_evo_floq}
\end{equation}

\noindent
Using the periodicity of the Floquet modes $ \phi_{\alpha} (t_0+T) =  \phi_{\alpha}  (t_0)$, Eq. (\ref{time_evo_floq}) can be reduced to the following eigenvalue problem

\begin{equation}
U \left(t_0 + T, t_0 \right)  \lvert \phi_{\alpha} (t_0) \rangle  = e^{\frac{-i \epsilon_{\alpha} T}{\hbar}}  \lvert \phi_{\alpha} (t_0 ) \rangle  .
\label{eig_time}
\end{equation}

\noindent 
If we take the initial time $t_0 = 0$, then the Floquet mode $ \lvert \phi_{\alpha} (0) \rangle $ and the quasienergies $\epsilon_{\alpha}$ can be found by finding the eigenfunctions and eigenvalues of $U \left(T, 0 \right)$. The Floquet modes $ \lvert \phi_{\alpha} (t) \rangle $ at any time instant $t$ are then obtained from the propagation of $ \lvert \phi_{\alpha} (0) \rangle $ with $U \left(t, 0 \right)$. Note that we only need to evaluate $ \lvert \phi_{\alpha} (t) \rangle $ from $0$ to $T$, as any other values of the Floquet modes at other times are fixed by their periodicity.

To summarize, in order to describe the time-evolution of a system we use the following approach: i. find the one-period time-evolution operator $U \left(T, 0 \right)$; ii. determine its eigenfunctions $ \lvert \phi_{\alpha} (0) \rangle $ and eigenvalues $e^{\frac{-i \epsilon_{\alpha} T}{\hbar}}$ by solving the eigenvalue problem (\ref{eig_time}); iii. find the decomposition of $\lvert \psi (0) \rangle$ in terms of Floquet modes: $c_{\alpha}  = \langle \phi_{\alpha} (0) \rvert \psi (0) \rangle$; iv. calculate the state of the system at time $t$ using the time-evolution operator $\lvert \psi (t) \rangle = \sum_{\alpha} c_{\alpha}e^{\frac{-i \epsilon_{\alpha} t}{\hbar}} U \left(t, 0 \right)  \lvert \phi_{\alpha} (0) \rangle$. Therefore, the use of the Floquet theorem allows us to reduce the problem of solving a differential equation into the problem of finding the eigenvalues and eigenfunctions of a matrix, which in some cases can be easier to solve.

\subsection{Analytical results}
\label{fl_an}
As a first step, we solve the problem analytically using the approach described in the previous Section. In general, the time-evolution operator $U(T , 0)$ can be obtained from the infinite series \cite{sakurai}

\begin{equation}
\begin{split}
\label{U_series}
U( T , 0) = 1 + \sum_{n=1}^{\infty}\frac{1}{n!} \left( -\frac{i}{\hbar} \right)^n \int_0^{T} dt^{(1)} \int_0^{T} dt^{(2)} ...  \int_0^{T} dt^{(n)} \hat{H} (t^{(1)}) \hat{H} (t^{(2)}) ...  \hat{H} (t^{(n)}) .
\end{split}
\end{equation}

\noindent
Since the Hamiltonian (\ref{H_qub_em}) is time-dependent and its terms do not all commute with each other, the analytical expression of the time-evolution operator (\ref{U_series}) for Hamiltonian (\ref{H_qub_em}) can only be found approximately. Using the Trotter-Suzuki formula \cite{trotter, suzuki1, suzuki2}, in the case of a time-dependent Hamiltonian \cite{poulin}, we write

\begin{equation}
\begin{split}
\label{U_trotter}
U( T , 0) \approx U \left( (N_t-1) \tau + \tau ,(N_t -1) \tau \right)...U (\tau + \tau , \tau) U (\tau , 0) ,
\end{split}
\end{equation}

\noindent
where the dynamics of the system is decomposed in a discrete number of steps $N_t$, with $\tau = T / N_t $ being the infinitesimal time-step of the Trotter decomposition. Eq. (\ref{U_trotter}) becomes exact as $N_t \rightarrow \infty$. At each step, the time-evolution operator can be factorized in two parts

\begin{equation}
\begin{split}
\label{U_trotter1}
U( \tau , 0) = U_0 ( \tau , 0) U_I ( \tau , 0) ,
\end{split}
\end{equation}

\noindent
where $U_0 ( \tau , 0) $ and $U_I ( \tau , 0) $ are the time-evolution operators corresponding to the Hamiltonian $\hat{H}_0$ and $\hat{H}_I$, respectively. Assuming the coupling to be "on" for half of the period $T$ with strength $g_0$, and "off" otherwise, as in Ref. \cite{lamata}, Eq. (\ref{U_trotter}) reduces to

\begin{equation}
\begin{split}
\label{U_trotter2}
U( T , 0)  \approx U_0 ( \tau , 0)^{N_t} U_I ( \tau , 0)^{\frac{N_t}{2}} =\left[   e^{ -i \omega _{r}\hat{a}^{\dagger }\hat{a} \tau} \prod_j^N \left\{ e^{-i \omega _{0}^j \hat{\sigma}^{+}_j \hat{\sigma}^{-}_j \tau } e^{-i \frac{g_0}{2} \hat{\sigma}_{x}^j \left( \hat{a}^{\dagger }+\hat{a}\right) \tau } \right\} \right]^{N_t}.
\end{split}
\end{equation}

\noindent
In particular, considering the case of a system of $N=1$ qubit coupled to a resonator with $n = 0,1$ photons we obtain the following matrix representation

\begin{equation}
U(T , 0) \approx 
\begin{pmatrix}
\cos \left(\frac{g_0 \tau}{2}\right) & 0 & 0 & -i \sin \left(\frac{g_0 \tau}{2}\right) e^{-i (\omega_0 + \omega_r) \tau} \\
0 & e^{-i  \omega_r \tau} \cos \left(\frac{g_0 \tau}{2}\right) & -i e^{-i  \omega_0 \tau} \sin \left(\frac{g_0 \tau}{2}\right) & 0\\
0 & -i e^{-i  \omega_r \tau} \sin \left(\frac{g_0 \tau}{2}\right) &e^{-i  \omega_0 \tau} \cos \left(\frac{g_0 \tau}{2}\right) & 0 \\
-i \sin \left(\frac{g_0 \tau}{2}\right) & 0 &0 & \cos \left(\frac{g_0 \tau}{2}\right) e^{-i  (\omega_0 + \omega_r) \tau}
\end{pmatrix}^{N_t}.
\label{U_matrix}
\end{equation}

\noindent
The Floquet approach can now be used to study the Trotterized dynamics of the system. As proven in the previous Section, the eigenvalues and the eigenvectors of the one-period time-evolution operator $U(T, 0)$ give an exponential function of the Floquet quasienergies and the Floquet modes, respectively. Their full expressions are given by Eqs. (\ref{quasien}) and (\ref{floq_mod}) in Appendix \ref{appA}. With these results, we can now find the wavefunction of the system at any time $t$ by following the procedure outlined in Sec. \ref{floquet}. First, we decompose the initial state of the system $\lvert \psi (0) \rangle $ in terms of Floquet modes at time $t=0$, then we propagate the initial state by using the one-period time-evolution operator. By noting that $U \left(T,0 \right) \approx U \left(\tau,0 \right) ^ {N_t}$, as proved in Ref. \cite{grifoni}, we can write the wavefunction as

\begin{equation}
\lvert \psi (T) \rangle  \approx \sum_{\alpha} c_\alpha (0)  U_0 ( \tau , 0)^{N_t} U_I ( \tau , 0)^{\frac{N_t}{2}} \lvert \phi_{\alpha} (0) \rangle e^{\frac{-i \epsilon_{\alpha} T }{\hbar}}.
\label{psi_t}
\end{equation}

\begin{figure}[]
	\subfloat[]{
	\includegraphics[width=7.3cm]{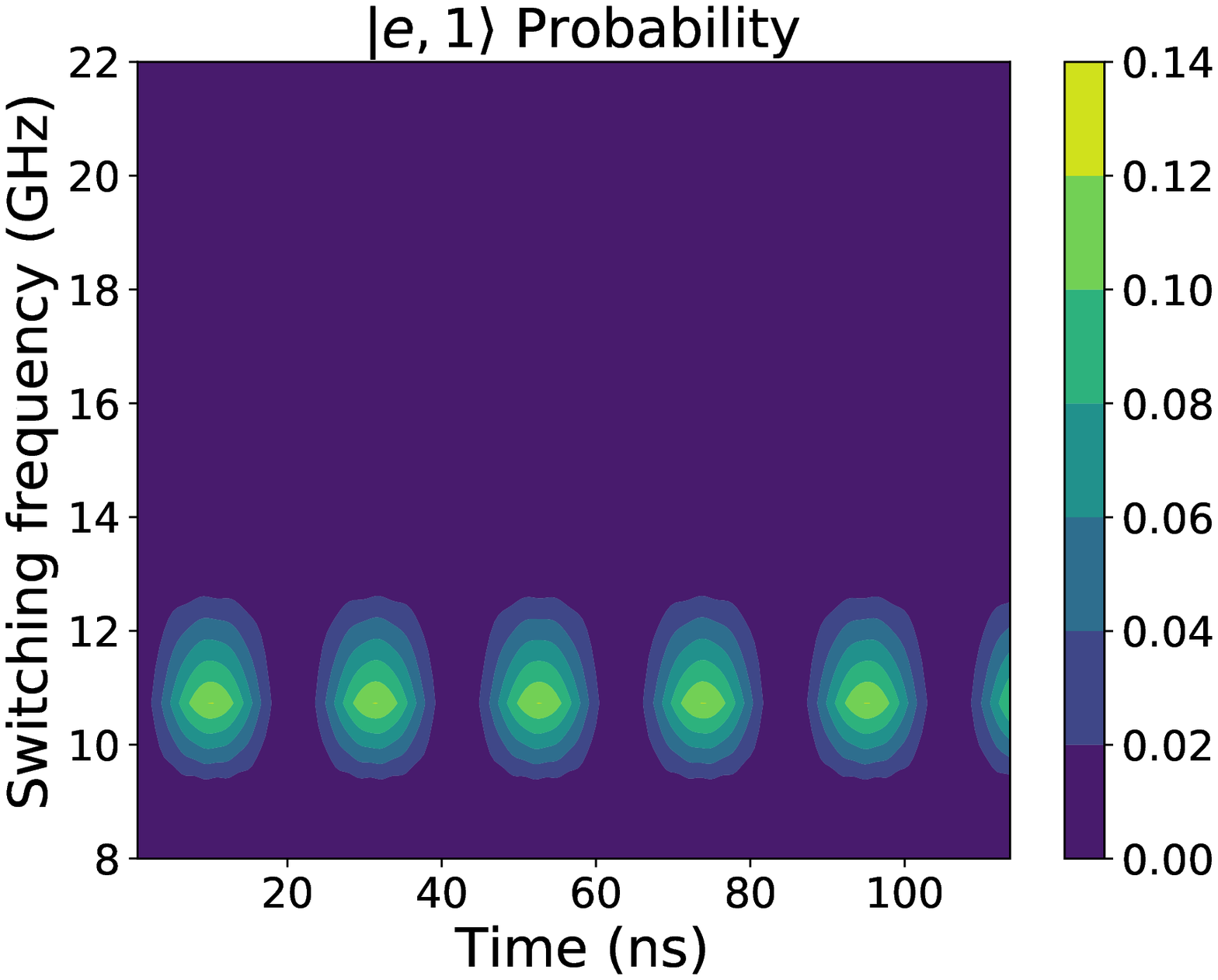}
	\label{2d_p_fl_an}
	}
	\subfloat[]{
	\includegraphics[width=7.3cm]{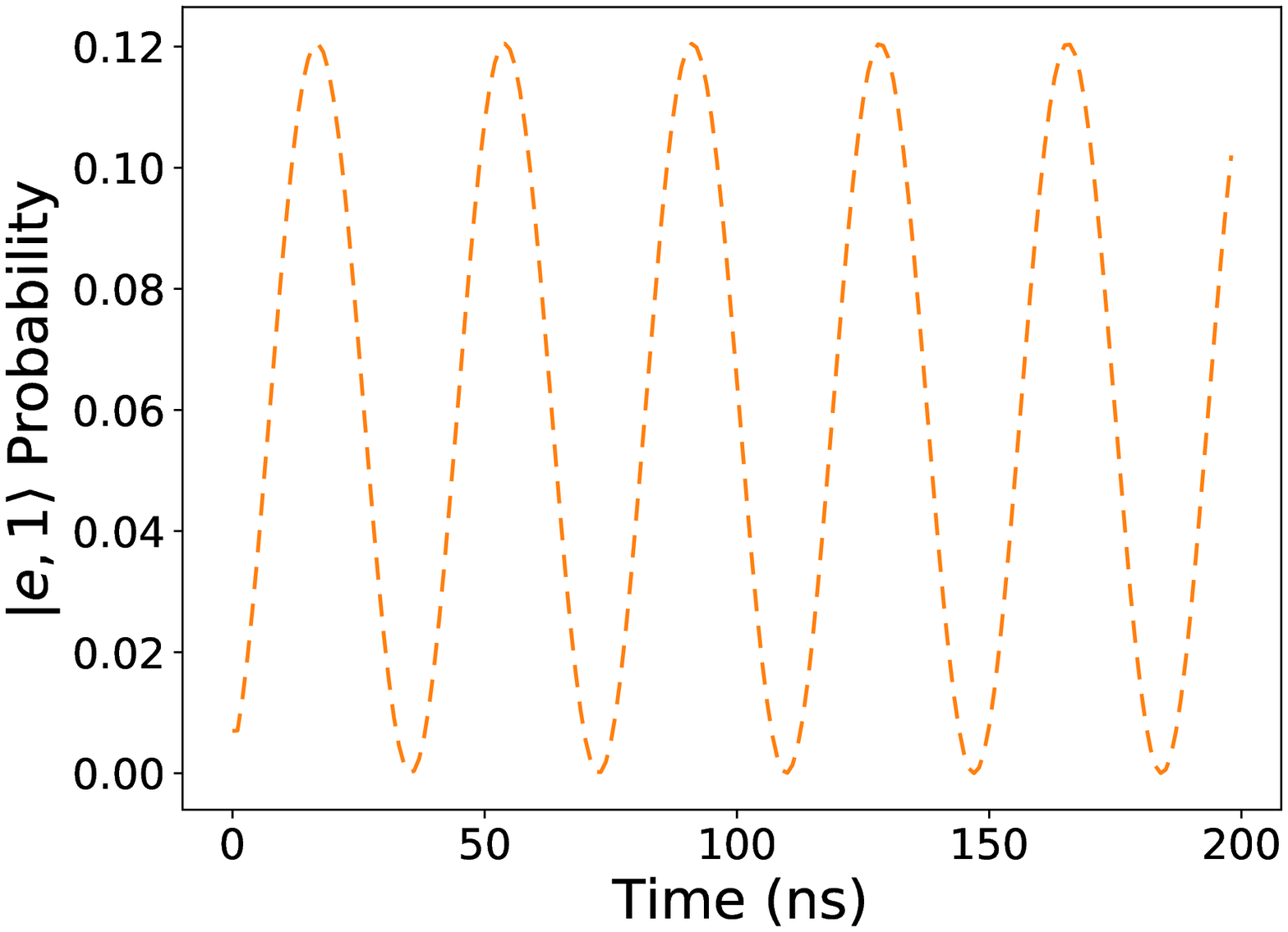}
	\label{1d_p_fl_an}
	} \\

\caption{(a) Two dimensional plot of the time-dependence of the probability of exciting the system to the $\lvert e, 1 \rangle $ state, where different values of the frequency of switching of the qubit/resonator coupling $\varpi_s$ are plotted on the y-axis. (b) Time-dependence at switching frequency $\varpi_s = \omega_0 + \omega_r$. }
\label{floquet_result_an}
\end{figure}

\noindent
The results of our calculations are presented in Fig. \ref{floquet_result_an}. Fig. \ref{2d_p_fl_an} shows the time dependence of the probability of finding the system in the $\lvert e, 1 \rangle $ state for different values of the switching frequency $\varpi_s$, in the case of a system initially in the ground state $\lvert \psi (0) \rangle = \lvert g, 0 \rangle $. The particular case where the switching frequency $\varpi_s$ equals the sum of the qubit and resonator frequencies $ \omega_0 + \omega_r$ is depicted in Fig. \ref{1d_p_fl_an}. When $\varpi_s$ takes this value, Fig. \ref{1d_p_fl_an} shows that the probability reaches its maximum. In this calculation, and throughout the rest of the paper, the following values where chosen for the parameters of the system: $\omega_0 = 5$ GHz, $\omega_r = 6$ GHz, $g_0 = 0.1$ GHz and $\varpi_s$ varies in the range $\left[ \frac{ \omega_0 + \omega_r}{2}, \frac{3  (\omega_0 + \omega_r)}{2} \right]$.

\begin{figure}[]
	\subfloat[]{
	\includegraphics[width=7.3cm]{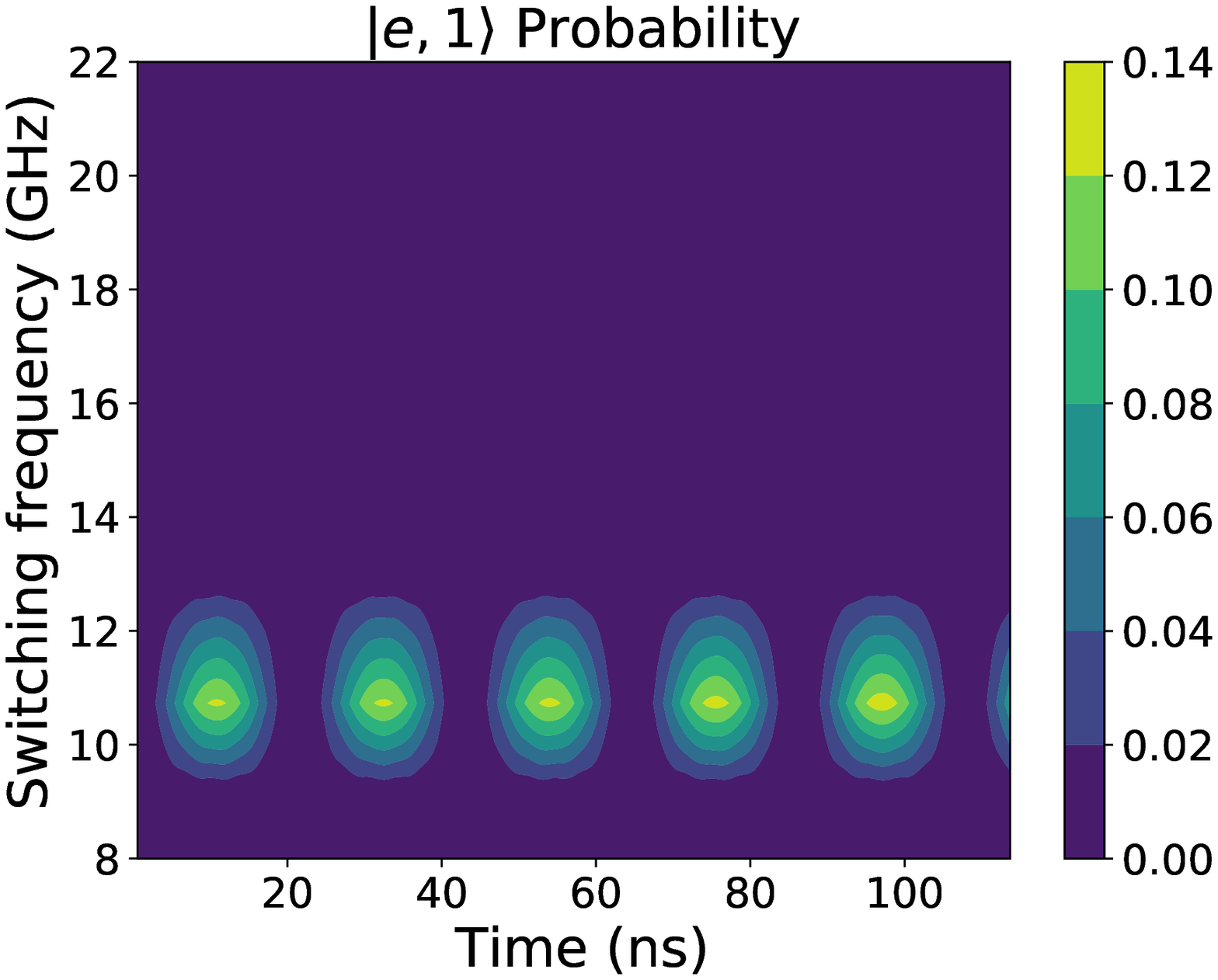}
	\label{2d_p_fl_num}
	}
	\subfloat[]{
	\includegraphics[width=7.3cm]{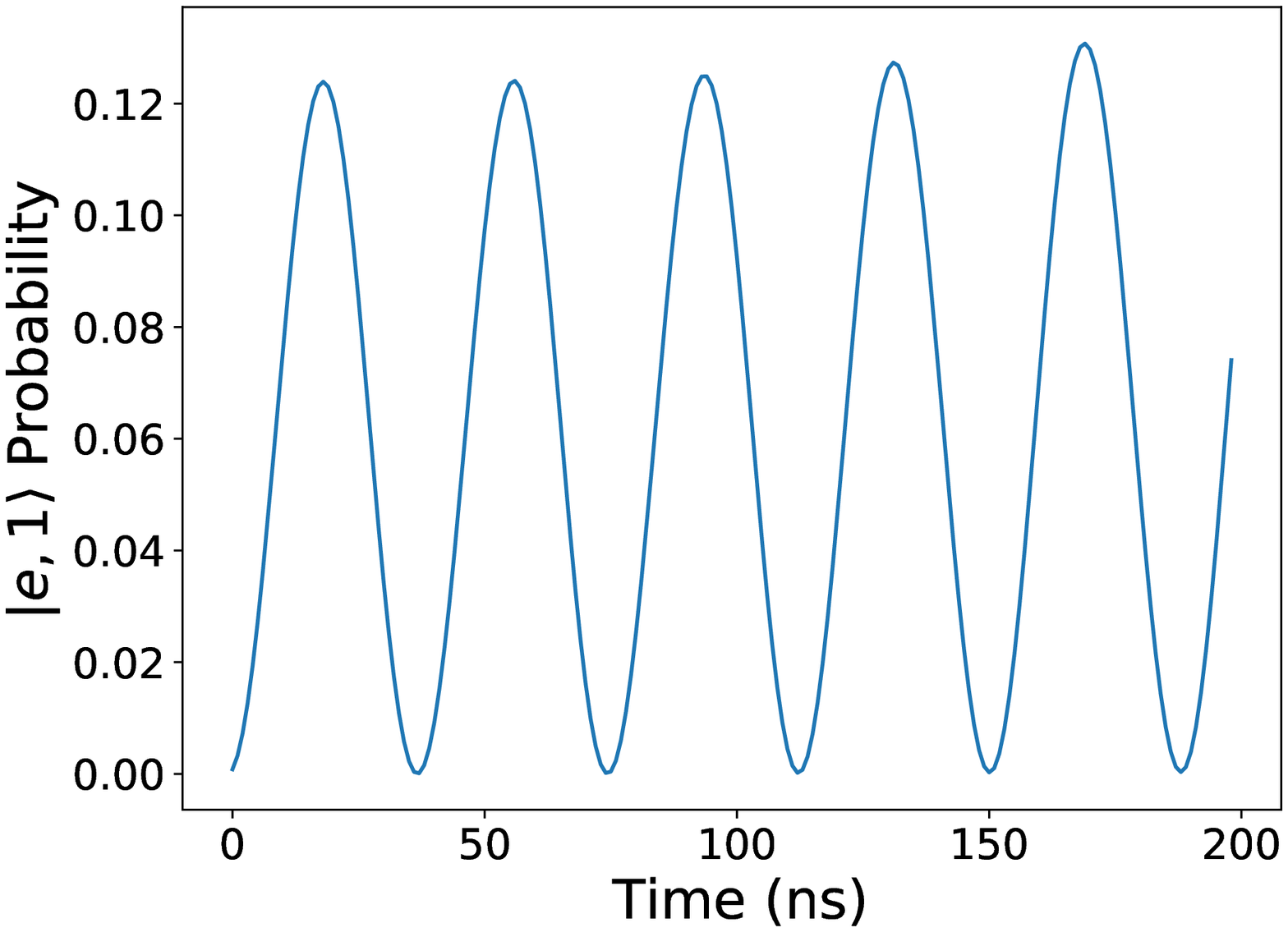}
	\label{1d_p_fl_num}
	} \\
\caption{ Numerical calculations of the time-dependence of the probability of finding the system in the $\lvert e, 1 \rangle $ state using the Floquet approach (a) contour-plot where different values of the frequency of switching of the qubit/resonator coupling $\varpi_s$ are plotted on the y-axis. (b) slice for the switching frequency $\varpi_s = \omega_0 + \omega_r$. }
\label{floquet_result_num}
\end{figure}

\begin{figure}[]
	\includegraphics[width=7.3cm]{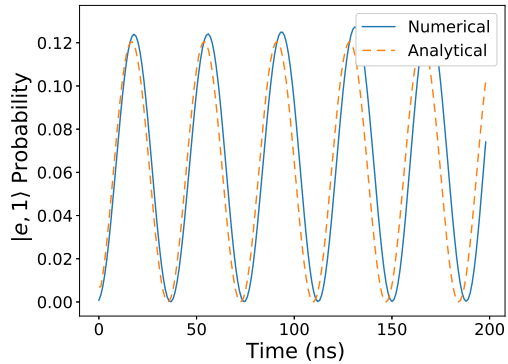}	
	\caption{Comparison of the time-dependence of the probability of exciting the $\lvert e, 1 \rangle $ state for the switching frequency $\varpi_s = \omega_0 + \omega_r$ calculated analytically and numerically. }
\label{floquet_result_num_an}
\end{figure}

The same procedure can be used for the numerical calculations. First, one finds the one-period time-evolution operator, then the eigenvalues of time-evolution of the system in the Floquet approach are also numerically calculated. The results obtained using the QuTip package \cite{qutip1, qutip2} are depicted in Fig. \ref{floquet_result_num}. Clearly the analytical and the numerical results are in good agreement with each other. A direct comparison of the two results can be seen in Fig. \ref{floquet_result_num_an}. 

%

%

%
%

\section{Comparison with other methods}
\label{comp}

The dynamics of the system can be studied by solving the time-dependent Schr\"{o}dinger equation within the framework of other approaches as well. Let us  show two of such approaches. First, we will solve the set of linear differential equations for the system using a perturbative approach. Second, we will use another equivalent approach which involves the Laplace transform. This can sometimes simplify the expression of time-dependent quantities, making it easier to describe the system's dynamics.

\subsection{Perturbative approach in the time-domain}
\label{pert}

The dynamical behavior of the system can be found by solving the Schr\"{o}dinger equation (\ref{schro}). For a system of $N=1$ qubit and $n=0,1$ photons described by Hamiltonian (\ref{H_qub_em}) we can write the wavefunction as 

\begin{equation}
\label{psi}
\lvert \psi \left( t \right)  \rangle  =  \alpha_{g, 0} \left( t \right) \lvert g, 0   \rangle + \alpha_{g, 1} \left( t \right) \lvert g, 1   \rangle+ \alpha_{e, 0} \left( t \right) \lvert e, 0   \rangle + \alpha_{e, 1} \left( t \right) \lvert e, 1   \rangle,
\end{equation}

\noindent
where indices $g$ and $e$ correspond to ground and excited state of the qubit, $0,1$ counts the number of photons and $\alpha(t)$ are the time-dependent amplitudes. The Schr\"{o}dinger equation then gives a set of coupled differential equations

\begin{eqnarray}
\label{schro_full}
i \frac{d {\alpha}_{g,0} (t) }{dt} &=& g(t) {\alpha}_{e,1} (t) , \nonumber \\ 
i \frac{d {\alpha}_{g,1} (t) }{dt} &=& \omega_r {\alpha}_{g,1} (t) + g(t) {\alpha}_{e,0} (t)  ,\nonumber \\
i \frac{d {\alpha}_{e,0} (t) }{dt} &=& \omega_0 {\alpha}_{e,0} (t) + g(t) {\alpha}_{g,1} (t)  ,\nonumber \\
i \frac{d {\alpha}_{e,1} (t) }{dt} &=&\Omega_+  {\alpha}_{e,1} (t) + g(t) {\alpha}_{e,0} (t)  .
\end{eqnarray}

\noindent
Here we have defined $\Omega_+ \equiv \omega_0 + \omega_r$ to simplify readability. One way to find a solution to the system of linear differential equations is to take a perturbative approach \cite{shapiro, zhukov, remizov, amico2}. Consider the following Hamiltonian  

\begin{equation}
\label{H}
\hat{H}\left( t \right) = \hat{H}_{0} +\delta \hat{H}_{I}\left( t \right) ,
\end{equation}

\noindent
where $\hat{H}_0$ is taken as the non-interacting Hamiltonian, $\hat{H}_I$ is taken as the perturbation and $\delta$ is a dimensionless parameter between 0 and 1. If the coupling strength $\delta g_0$ between the qubit and the resonator is small compared to the spacing of the energy levels of the unperturbed Hamiltonian $\hat{H}_0$, then the wavefunction can be expanded in powers of $\delta$

\begin{equation}
\label{psi_exp}
{\lvert \psi \left( t \right)  \rangle} = {\lvert \psi \left( t \right)  \rangle}^{(0)} +\delta {\lvert \psi \left( t \right)  \rangle}^{(1)} + \delta^2 {\lvert \psi \left( t \right)  \rangle}^{(2)} + ... \, .
\end{equation}

%
%
%
%

\noindent
As a result one obtains the following differential equations for any other order $(j)$ of the perturbation

\begin{equation}
\label{schro_j}
i \frac{d {\lvert \psi \left( t \right)  \rangle}^{(j)} }{dt} = \hat{H}_0  {\lvert \psi \left( t \right)  \rangle}^{(j)} +  \hat{H}_{I}\left( t \right)  {\lvert \psi \left( t \right)  \rangle}^{(j-1)} .
\end{equation}

\noindent
One can then solve the Schr\"{o}dinger equation order by order. For the case considered here, an analytical solution to the Schr\"{o}dinger equation up to second order in the perturbation $\delta$ is found. The calculation of the values of the coefficients of the wavefunction (\ref{psi}) is presented in Appendix \ref{appB}. The resulting wavefunction reads

\begin{eqnarray}
\label{psi_2_pert}
\begin{split}
{\lvert \psi \left( t \right)  \rangle} =  {\lvert g,0  \rangle}^{(0)} - \delta g_0 e^{-i \Omega_+t } \frac{ \left\{  \Omega_+ + e^{i   \Omega_+ t} \left[ i \varpi_s \sin \left( \varpi_s t \right) -   \Omega_+ \cos \left( \varpi_s t \right) \right] \right\} }{\left[ \varpi_s +  \Omega_+ \right] \left[ \varpi_s -  \Omega_+ \right] } {\lvert e,1  \rangle}^{(1)} + \\ 
+ \left(\delta g_0\right)^2 \frac{ \left\{  i \varpi_s \left[ -1 + 2 i  \Omega_+ t +\cos  \left( 2 \varpi_s t \right)  \right] -   \Omega_+ \sin  \left( 2 \varpi_s t \right)  \right\} }{ 4 \varpi_s \left[ \varpi_s +  \Omega_+ \right] \left[ \varpi_s -  \Omega_+ \right]  } {\lvert g,0  \rangle}^{(2)} .
\end{split}
\end{eqnarray}

\begin{figure}[]
	\subfloat[]{
	\includegraphics[width=7.3cm]{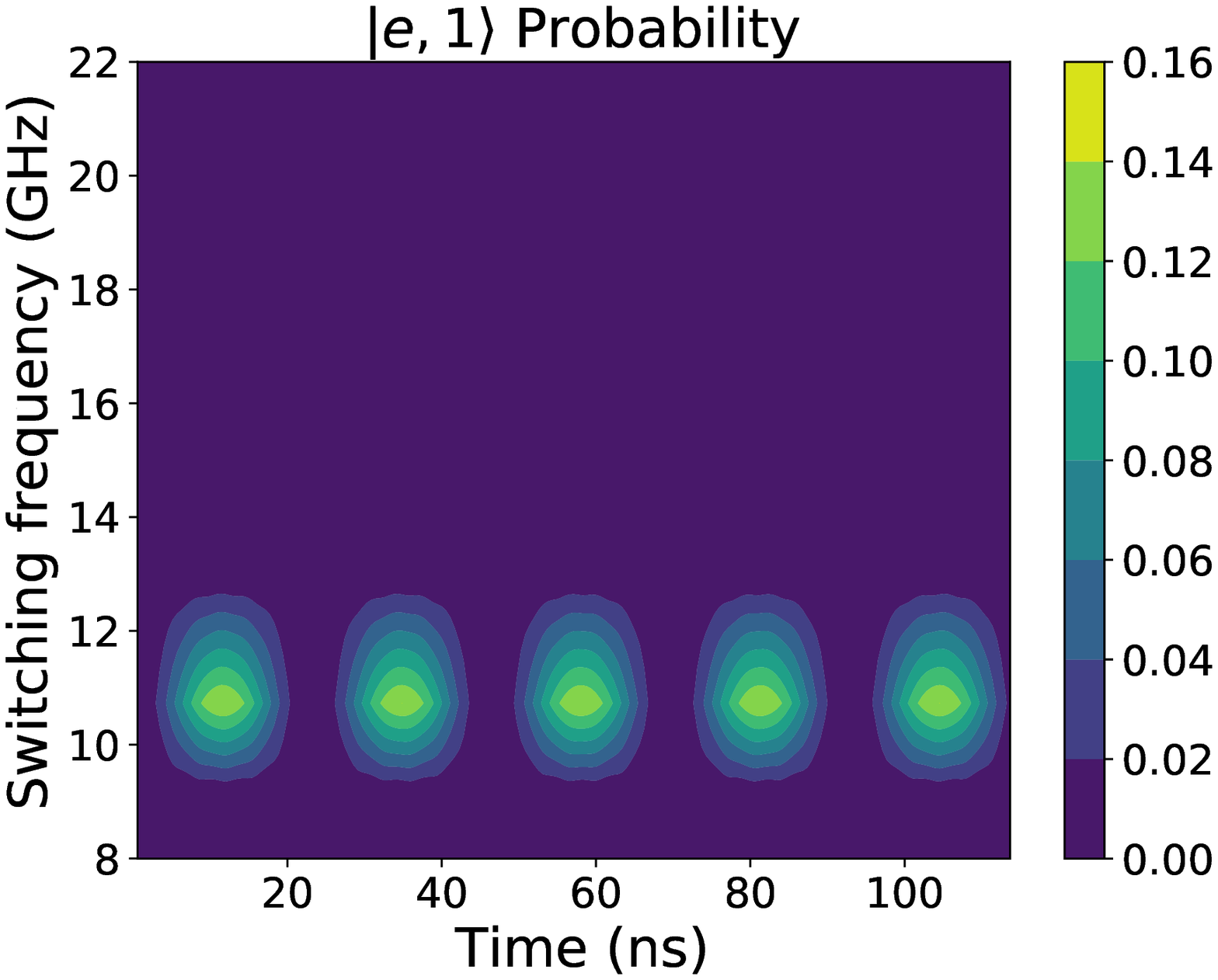}
	\label{2d_p_s_an}
	}
	\subfloat[]{
	\includegraphics[width=7.3cm]{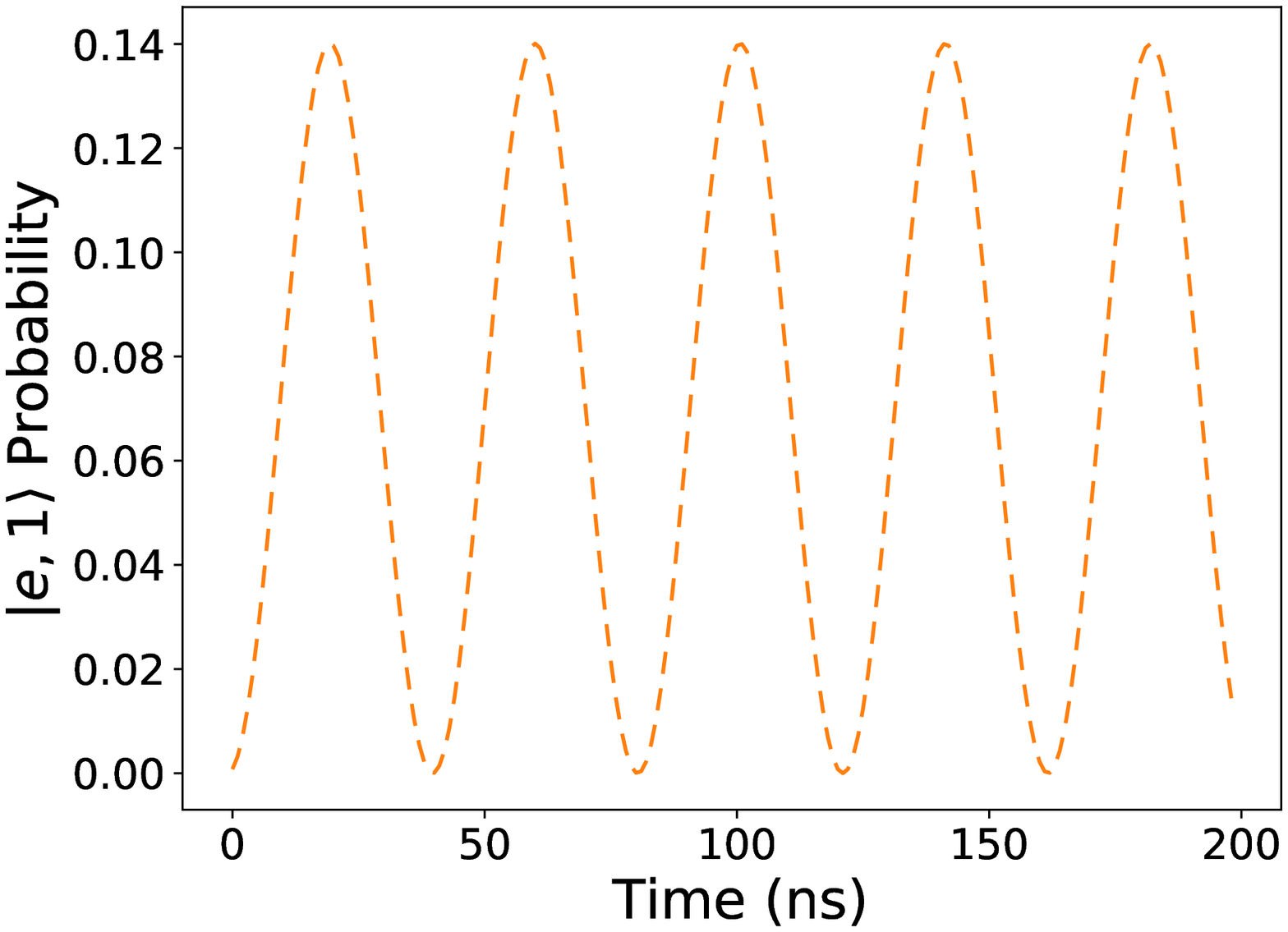}
	\label{1d_p_s_an}  
	} \\
	\subfloat[]{
	\includegraphics[width=7.3cm]{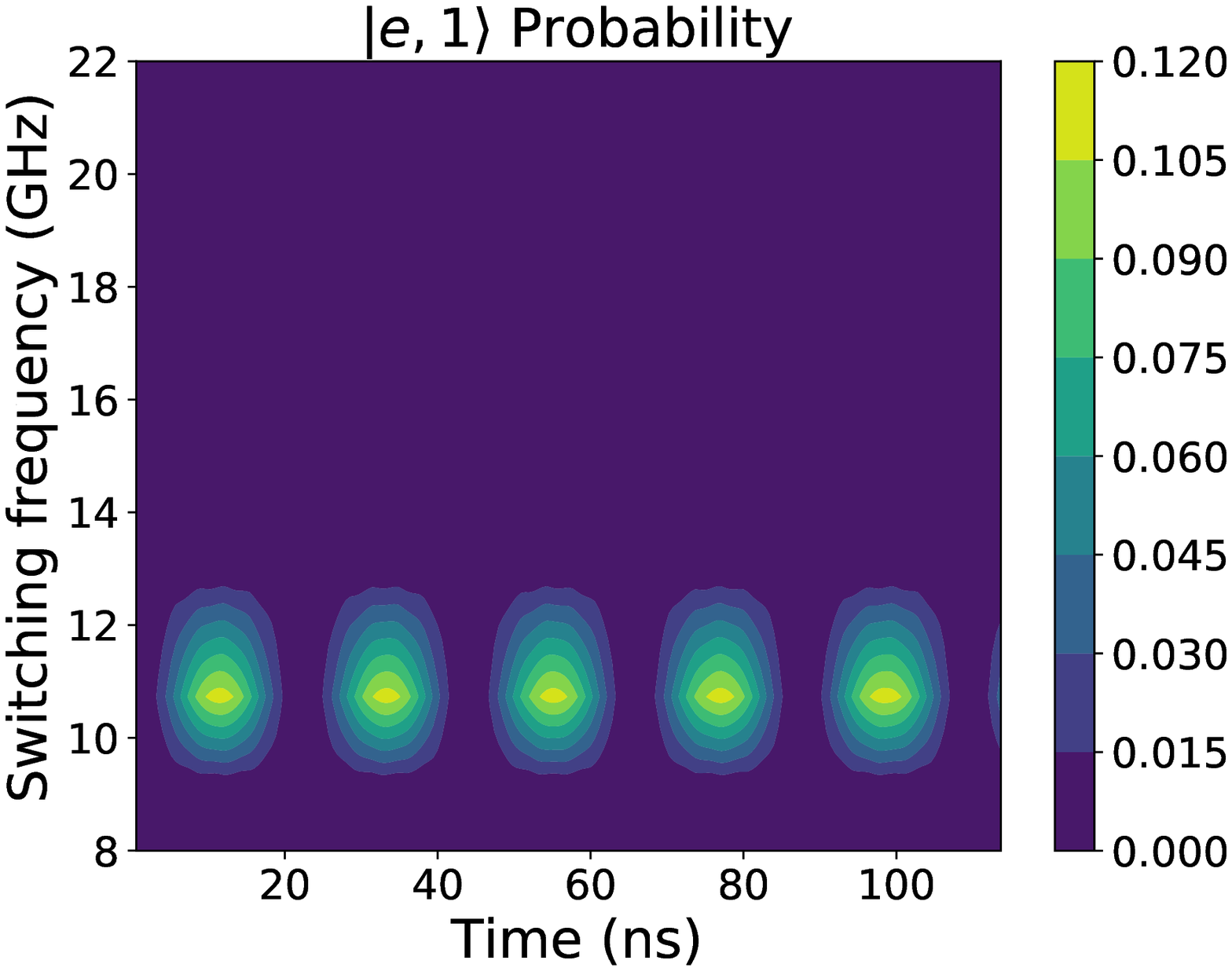}
	\label{2d_p_s_num}
	}
	\subfloat[]{
	\includegraphics[width=7.3cm]{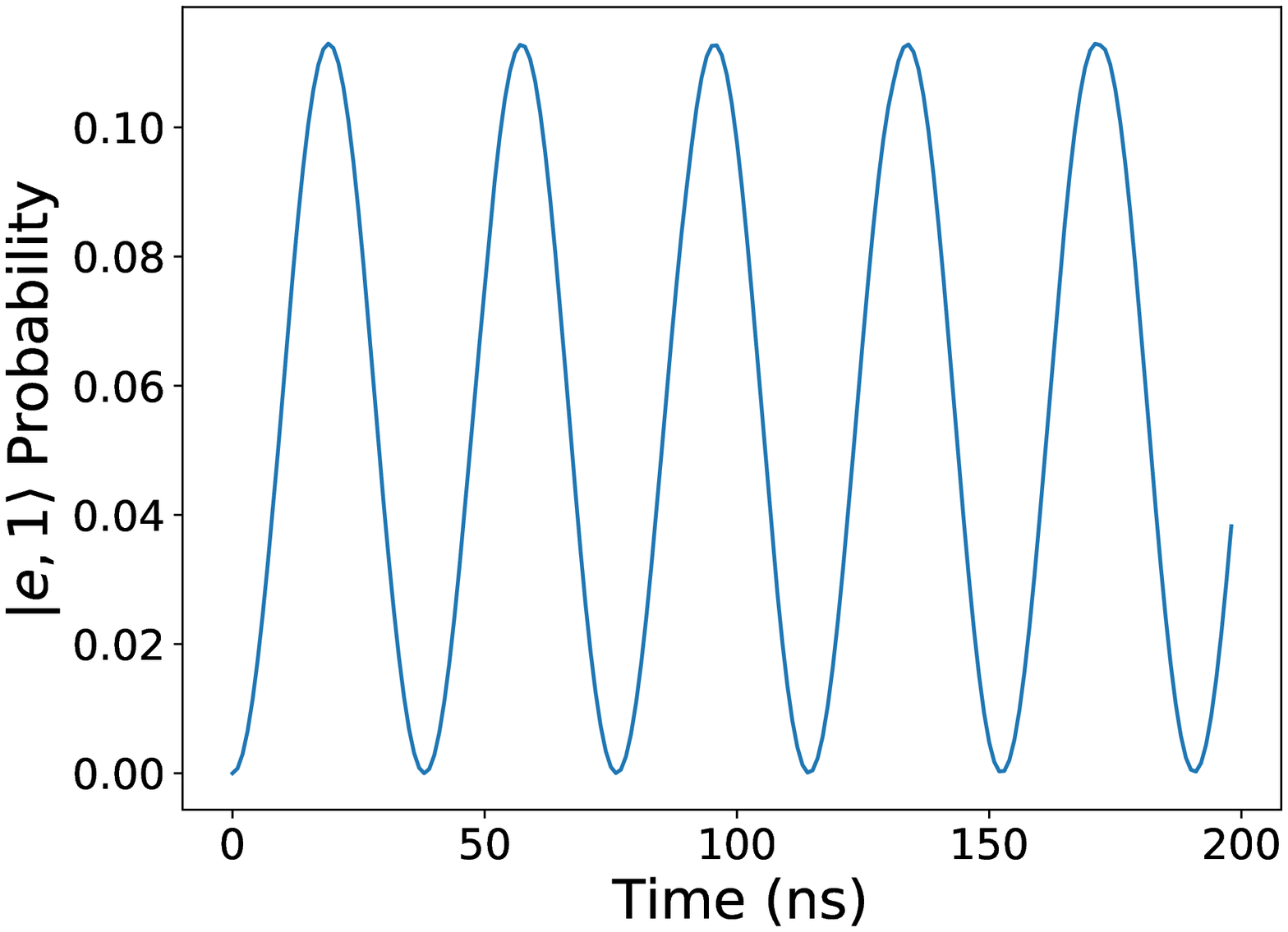}
	\label{1d_p_s_num}  
	} \\	
\caption{ The time-dependence of the probability of finding the system in the $\lvert e, 1 \rangle $ state is determined using a perturbative approach. The results in (a) and (b) are obtained by analytically solving the Schr\"{o}dinger equation within a perturbative approach. (c) and (d) show the results obtained by numerical integration of Schr\"{o}dinger's equation. (a), (c) show a two-dimensional plot for different values of the frequency of switching of the qubit/resonator coupling $\varpi_s$. (b), (d) show a one-dimensional plot for $\varpi_s =\Omega_+$. }
\label{perturbative_result_an}
\end{figure}

\noindent
Using the expression above, one can study the dynamics of the system. In Figs. \ref{2d_p_s_an} and \ref{1d_p_s_an}, the time-evolution of the probability of finding the system in the state $ \lvert e, 1 \rangle $ using the perturbative solution (\ref{psi_2_pert}) is presented along with the results obtained from numerical integration of the Schr\"{o}dinger equation shown in Figs.  \ref{2d_p_s_num} and \ref{1d_p_s_num}.

\begin{figure}[]
	\includegraphics[width=7.3cm]{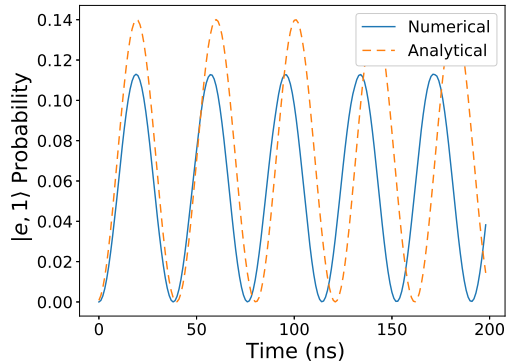}	
	\caption{Comparison of the analytical solution of the Schr\"{o}dinger equation within a perturbative approach (dashed curve) with 
the numerical integration of this equation (solid curve). }
\label{perturbative_result_num_an}
\end{figure}

\noindent
A direct comparison between the time-evolution of the probability that the system is in the $\lvert e, 1 \rangle $ state obtained using the analytical solution found via the perturbative approach and the numerical solution of the Schr\"{o}dinger equation is shown in Fig. \ref{perturbative_result_num_an}.

\subsection{Perturbative approach in the Laplace-domain}
\label{lap}

The perturbative approach in the Laplace-domain presents an alternative approach that can be used when time-dependent parameters make the direct solution of the Schr\"{o}dinger equation too difficult. Similarly to the Fourier transform, the Laplace transform can be used to reduce the problem at hand into a different problem, which is sometimes easier to solve. In fact, a set of linear differential equations in the time-domain can be turned into a set of algebraic equations in the Laplace-domain. The price to pay is the cost of the transformation from the time-domain to the Laplace-domain and back. The Laplace transform of a function $f (t)$ is defined as

 \begin{equation}
 F (s) = \mathcal{L}\left[ f (t) \right](s) = \int_0^{\infty} dt \, f (t) e^{-st} ,
\end{equation}

\noindent
where $s$ is a complex number. Thus the Laplace transform requires us to compute an integral. On the other hand, the possibility of using Cauchy's residue theorem reduces the calculation of the inverse Laplace transform to the calculation of \textit{Residues}

\begin{equation}
\label{inv_laplace} 
\mathcal{L}^{-1}\left[ F(s) \right](t) = \frac{1}{2\pi i} \int_{b-i\infty}^{b+i\infty} ds \, F(s) e^{st}  = \sum_{\text{poles of } F(s)} \text{Res}\left( F(s)e^{st} \right), 
\end{equation}

\noindent
where $b$ is a point on the real axis on the right of the rightmost pole of $F(s)$.
To find the dynamics of the system considered above within this approach, one starts from the Hamiltonian (\ref{H}) and the perturbative expansion of the wavefunction (\ref{psi_exp}). Then applying the Laplace transform to the set of linear differential equations (\ref{coeff1q_0_pert}), (\ref{coeff1q_1_pert}) and (\ref{coeff1q_2_pert}) given in Appendix \ref{appB}, one can solve the algebraic system of equation order by order. A solution in the time-domain is then obtained by taking the inverse Laplace transform. Using the results obtained in Appendix \ref{appC}, one can write the approximate solution of the Schr\"{o}dinger equation within the Laplace approach as

\begin{eqnarray}
\label{psi_2_lap}
\begin{split}
{\lvert \psi \left( t \right)  \rangle} =  {\lvert g,0  \rangle}^{(0)} - \frac{\delta g_0}{2} \left[  \frac{e^{i \varpi_s t} }{\varpi_s -  \Omega_+ } - \frac{e^{i \varpi_s t} }{\varpi_s +  \Omega_+ }    - \frac{2 e^{- i  \Omega_+  t}  \Omega_+  }{\varpi_s^2 -  \Omega_+^2 } \right] {\lvert e,1  \rangle}^{(1)} + \\ 
+ \frac{1}{16} \left( \delta g_0 \right)^2 \left(\frac{\varpi_s ^2 (6-4 i \Omega_+ t)+2 i \Omega_+^2 (2 \Omega_+  t +7 i)}{(\varpi_s -\Omega_+)^2 (\varpi_s +\Omega_+)^2}+\frac{e^{-2 i \varpi_s t }}{\varpi_s(\varpi_s - \Omega_+)}+\frac{e^{2 i \varpi_s t}}{\varpi_s(\varpi_s +\Omega_+)}\right) {\lvert g,0  \rangle}^{(2)} .
\end{split}
\end{eqnarray}

\begin{figure}[]
	\subfloat[]{
	\includegraphics[width=7.3cm]{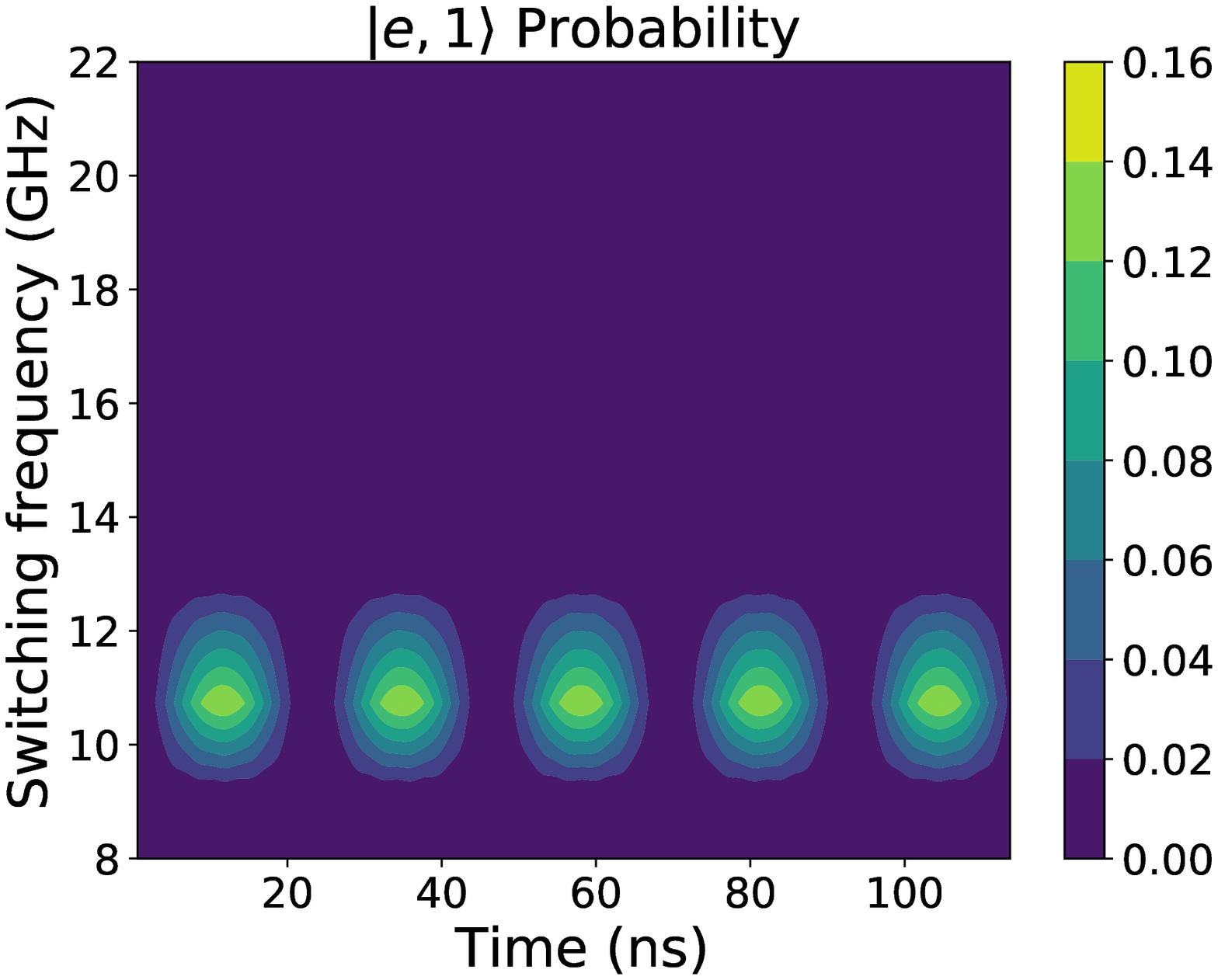}
	\label{2d_p_l_an}
	}
	\subfloat[]{
	\includegraphics[width=7.3cm]{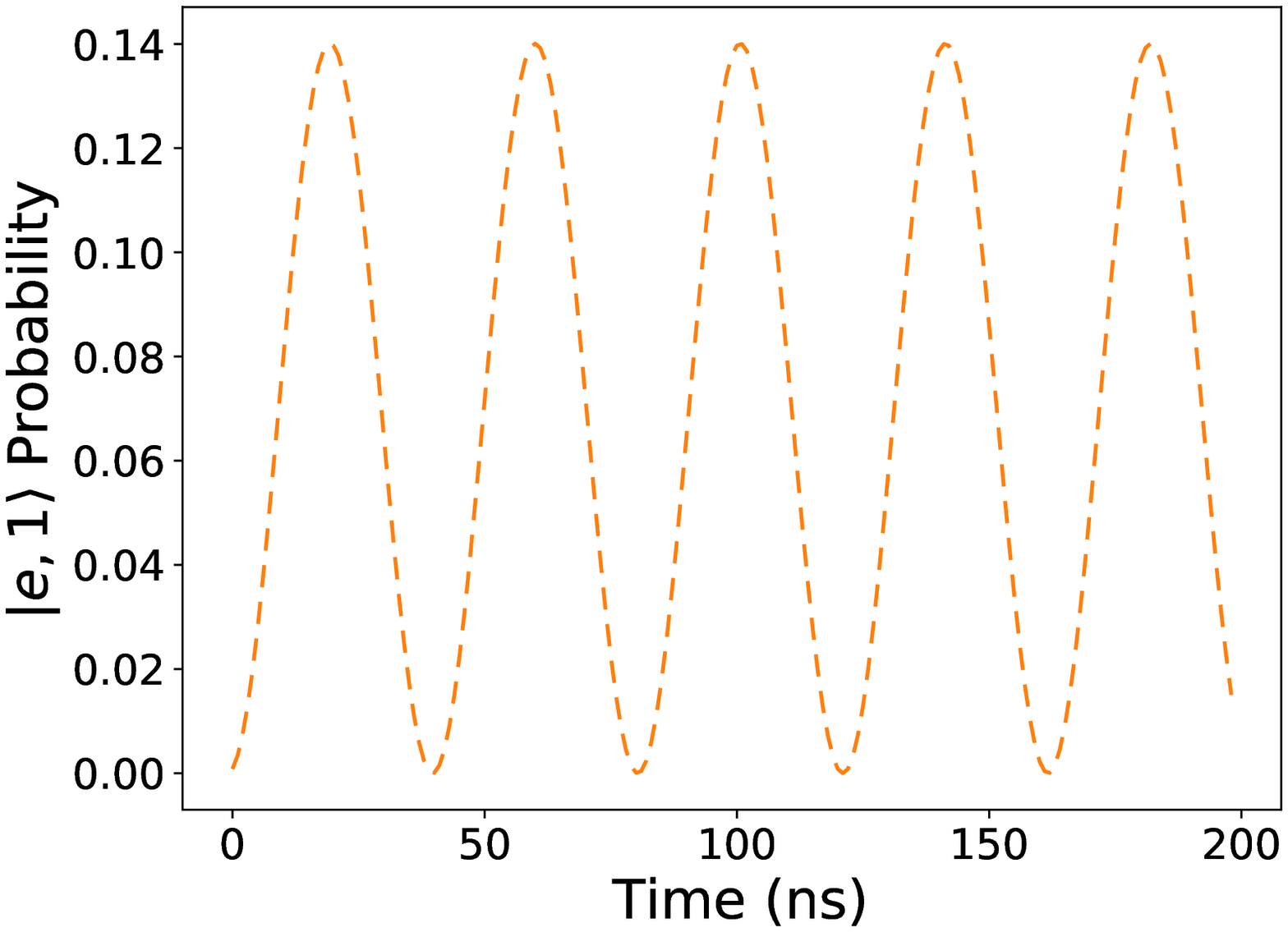}
	\label{1d_p_l_an}  
	} \\
\caption{Results for the time-dependence of the probability of finding the system in the $\lvert e, 1 \rangle $ state found using the Laplace approach. The analytical solution in Eq. (\ref{psi_2_lap}) is used to make the plots. (a) is a two-dimensional plot where different frequencies of switching of the coupling  $\varpi_s $ are considered and (b) is a one-dimensional slice for $\varpi_s = \Omega_+$. }
\label{laplace_result_an}
\end{figure}

\noindent
Results of the calculations for the time-dependence of the probability of finding the system in the $\lvert e, 1\rangle$ state found in the framework of the Laplace approach are presented in Fig. \ref{laplace_result_an}.


\section{Results and discussion}
\label{res}

\begin{figure}[]
	\includegraphics[width=10.3cm]{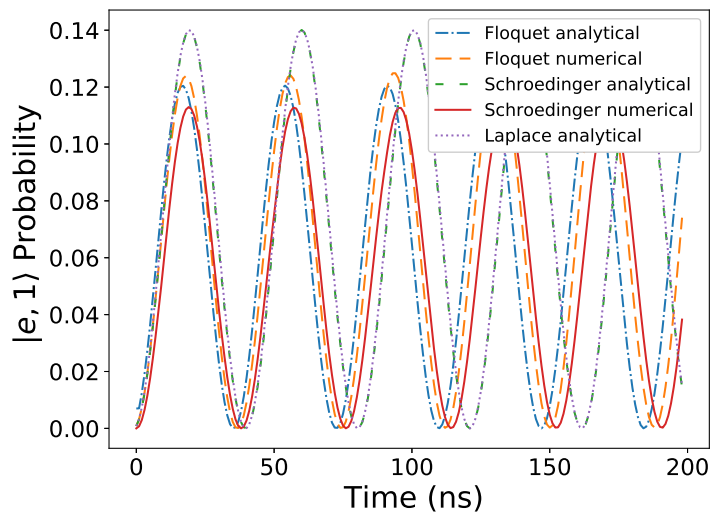}	
	\caption{Comparison of the results obtained using all methods presented in the paper. }
\label{final_comp}
\end{figure}

We presented three approaches for the description of the dynamics of periodically driven quantum system of $N$ qubits coupled to a resonator. A final comparison of all methods used to compute the dynamics of the $\lvert e, 1 \rangle $ state of the system is shown in Fig. \ref{final_comp}. All of them give similar results, although the analytical solutions tend to be accurate only at short times and then start to diverge from the numerical solutions. This is to be expected because the analytical solutions of the problem that we have found were obtained by taking some approximations. In Sec. \ref{fl_an} we used Trotterization to obtain an analytical expression of the time-evolution operator needed for the Floquet approach, in Sec. \ref{pert} and \ref{lap} perturbation theory was used and the infinite series was truncated at the second order in the perturbation parameter. Nonetheless, all qualitative features of the numerically solved solutions are displayed in the analytical solutions as well. Furthermore, the quantitative values of all solutions are quite close to each other.

\section{Conclusions}
\label{conc}

In conclusion, we have used the Floquet approach to give an outline of an analytical solution for the dynamics of a system of $N$ qubits coupled to a resonator with a periodically varying coupling. The case of a single qubit coupled to a resonator populated with $n=0,1$ photons was explicitly treated. The analytical solution of the Floquet problem found in this way closely resembles the solution that can be found numerically at short times. At long times, the two solutions differ noticeably but retain the same qualitative features. 

We have also studied the same problem adopting different analytical and numerical approaches. Whenever the Hamiltonian contains terms that can be considered as a perturbation, that is their effect is small compared to other terms in the Hamiltonian, an analytical solution to the Schr\"{o}dinger equation can be found using a perturbative approach. The latter can, in principle, be used to compute an approximate to any given order of the perturbative parameter. In the case considered here, such approach gives a solution that compares well with direct numerical integration. Moreover, the Laplace transform can be used to solve the Schr\"{o}dinger equation when particular time-dependence in the parameters of the system make an analytical approach in the time-domain difficult. An analytical solution based on such approach was also found. As expected, the results coincide with the ones obtained by solving Schr\"{o}dinger equation in the time-domain.

Overall, we showed that different analytical and numerical methods can be used to study the time-evolution of quantum systems. All analytical methods investigated give a good description of the dynamics of the system at short time, but a non-negligible discrepancy with the numerical methods arises at long times. Furthermore, it seems that the Floquet approach provides a framework where analytical and numerical methods give comparable results despite the approximations needed to obtain an analytical solution of the problem. We deduce that the Floquet method is the best method to use when dealing with time-periodic Hamiltonians. Nonetheless, the perturbative method in the time- and Laplace domain can be effectively used when aperiodic time-dependent terms appear in the Hamiltonian. In the latter case, the Laplace method is useful to deal with complicated time-dependencies while the direct integration of Schr\"{o}dinger's equation can only be done when simple time-dependent terms are present.

\acknowledgments
The authors are grateful to A. Kuklov for the valuable and stimulating discussions. This work is partially supported by the U.S. Department of Defense under Grant No. W911NF1810433.

\appendix
\section{}
\label{appA}

The quasienergies and the Floquet modes of the Hamiltonian (\ref{H_qub_em}) for the case of a single qubit coupled to a resonator with $n=0,1$ photons can be found by calculating the eigenvalues and eigenvectors of $U(\tau,0)$, the one-step time-evolution operator (\ref{U_matrix}). We get the following eigenvalues

\begin{equation}
\begin{aligned}
\label{quasien}
e^{\frac{-i \epsilon_{1} \tau}{\hbar}}   &= \frac{1}{2}  \cos \left(\frac{g\tau}{2}\right)  e^{-i \Omega_+ \tau} \left[ e^{ i \Omega_+ \tau}+1  - i \sqrt{ \left(e^{ i \Omega_+ \tau}+1\right)^2+4 \sec^2 \left(\frac{g\tau}{2}\right) e^{ i \Omega_+ \tau}}\right], \\
 e^{\frac{-i \epsilon_{2} \tau}{\hbar}} &= \frac{1}{2} \cos \left(\frac{g\tau}{2}\right)  e^{-i \Omega_+ \tau} \left[ e^{ i \Omega_+ \tau}+1 + i \sqrt{ \left(e^{ i \Omega_+ \tau}+1\right)^2+4 \sec^2 \left(\frac{g\tau}{2}\right) e^{ i \Omega_+ \tau}}\right], \\
e^{\frac{-i \epsilon_{3} \tau}{\hbar}} &= \frac{1}{2} \cos \left(\frac{g\tau}{2}\right)  e^{-i \Omega_+ \tau} \left[ e^{ i \Omega_+ \tau}+1  - i \sqrt{  \left(e^{ i \Omega_+ \tau}+e^{i \omega_r \tau} \right)^2 + 4 \sec^2 \left(\frac{g\tau}{2}\right) e^{ i \Omega_+ \tau}}\right], \\
e^{\frac{-i \epsilon_{4} \tau}{\hbar}} &=  \frac{1}{2} \cos \left(\frac{g\tau}{2}\right)  e^{-i \Omega_+ \tau} \left[ e^{ i \Omega_+ \tau}+1  + i \sqrt{  \left(e^{ i \Omega_+ \tau}+e^{i \omega_r \tau} \right)^2 + 4 \sec^2 \left(\frac{g\tau}{2}\right) e^{ i \Omega_+ \tau}}\right]  .
\end{aligned}
\end{equation}

\noindent
While for the eigenvectors we get

\begin{equation}
\label{floq_mod}
\begin{split}
\lvert \phi_{1} (0) \rangle   &=   \left\{ \frac{1}{2} i \cot \left(\frac{g\tau}{2}\right)  \left[ 1- e^{-i \Omega_+ \tau}   - e^{-i \Omega_+ \tau}  \sqrt{1+  2 e^{ i \Omega_+ \tau}\left( 1 - 2 \sec^2 \left(\frac{g\tau}{2}\right) \right) + e^{2 i \Omega_+ \tau}  }   \right],0,0,1\right\} , \nonumber  \\
\lvert \phi_{2} (0) \rangle   &=   \left\{ \frac{1}{2} i \cot \left(\frac{g\tau}{2}\right)  \left[ 1- e^{-i \Omega_+ \tau}   + e^{-i \Omega_+ \tau}  \sqrt{1+  2 e^{ i \Omega_+ \tau}\left( 1 - 2 \sec^2 \left(\frac{g\tau}{2}\right) \right) + e^{2 i \Omega_+ \tau}  }   \right],0,0,1\right\} ,  \nonumber  \\
\lvert \phi_{3} (0) \rangle   &=  \left\{0,\frac{1}{2} i \cot \left(\frac{g\tau}{2}\right)  e^{-i  \omega_0 \tau} \left[ e^{i \Omega_+ \tau} - e^{i \omega_r \tau} - \sqrt{ e^{2 i \omega_0 \tau} + e^{2 i \omega_r \tau} + 2 e^{i \Omega_+ \tau} \left(1 - 2 \sec^2 \left(\frac{g\tau}{2}\right) \right) }\right],1,0\right\} , \nonumber  \\
\lvert \phi_{4} (0) \rangle   &=  \left\{0,\frac{1}{2} i \cot \left(\frac{g\tau}{2}\right)  e^{-i  \omega_0 \tau} \left[ e^{i \Omega_+ \tau} - e^{i \omega_r \tau} + \sqrt{ e^{2 i \omega_0 \tau} + e^{2 i \omega_r \tau} + 2 e^{i \Omega_+ \tau} \left(1 - 2 \sec^2 \left(\frac{g\tau}{2}\right) \right) }\right],1,0\right\} .
\end{split}
\end{equation}

\section{}
\label{appB}

The perturbative approach allows one to rewrite the system of coupled differential equations (\ref{schro_full}) at each order of $\delta$. For the case of a system with the wavefunction (\ref{psi}), at the zero-th order in terms of $\delta$, Eq. (\ref{schro_j}) gives

\begin{eqnarray}
\label{coeff1q_0_pert}
i \frac{d {\alpha}^{(0)}_{g,0} (t)}{dt} &= & 0 ,\nonumber \\ 
i \frac{d {\alpha}^{(0)}_{g,1} (t)}{dt} &= & \omega_r {\alpha}^{(0)}_{g,1} (t) ,\nonumber \\
i \frac{d {\alpha}^{(0)}_{e,0} (t)}{dt} &= & \omega_0 {\alpha}^{(0)}_{e,0} (t),\nonumber \\
i \frac{d {\alpha}^{(0)}_{e,1} (t)}{dt} &= &   \Omega_+ {\alpha}^{(0)}_{e,1}(t)  .
\end{eqnarray}

\noindent
Given that the system is initially in the ground state $\lvert \psi (0) \rangle =   \lvert g, 0   \rangle $, thus the only non-zero coefficient is $ \alpha_{g, 0} \left( 0 \right) = 1$, one finds that ${\alpha}^{(0)}_{g,0} (t) = 1$ and ${\alpha}^{(0)}_{g,1} (t) = {\alpha}^{(0)}_{e,0} (t) = {\alpha}^{(0)}_{e,1} (t) = 0$.

\noindent
At first order in terms of $\delta$ one finds 
 
\begin{eqnarray}
\label{coeff1q_1_pert}
i \frac{d {\alpha}^{(1)}_{g,0} (t)}{dt} &= & g(t) {\alpha}^{(0)}_{e,1} (t) ,\nonumber \\ 
i \frac{d {\alpha}^{(1)}_{g,1} (t)}{dt} &= & \omega_r {\alpha}^{(1)}_{g,1}  (t) + g(t) {\alpha}^{(0)}_{e,0} (t) ,\nonumber \\
i \frac{d {\alpha}^{(1)}_{e,0} (t)}{dt} &= & \omega_0 {\alpha}^{(1)}_{e,0} (t) + g(t)  {\alpha}^{(0)}_{g,1} (t) ,\nonumber \\
i \frac{d {\alpha}^{(1)}_{e,1} (t)}{dt} &= &   \Omega_+ {\alpha}^{(1)}_{e,1} (t) +g(t)  {\alpha}^{(0)}_{g,0} (t).
\end{eqnarray}

\noindent
Substituting the value for the zero-th order coefficients $\alpha^{(0)} (t)$, one can solve for the first order coefficients. The only non-zero coefficients at first order is

\begin{equation}
\label{coef1q1sol_pert}
 {\alpha}^{(1)}_{e,1} \left( t \right) = - g_0 e^{-i  \Omega_+ t } \frac{ \left\{ \Omega_+ + e^{i   \Omega_+ t} \left[ i \varpi_s \sin \left( \varpi_s t \right) -   \Omega_+ \cos \left( \varpi_s t \right) \right] \right\} }{\left[ \varpi_s +  \Omega_+ \right] \left[ \varpi_s -  \Omega_+ \right] }.
\end{equation}

\noindent
At second order in terms of $\delta$, we have

\begin{eqnarray}
\label{coeff1q_2_pert}
i \frac{d {\alpha}^{(2)}_{g,0} (t)}{dt} &= & g(t) {\alpha}^{(1)}_{e,1} (t) ,\nonumber \\ 
i \frac{d {\alpha}^{(2)}_{g,1} (t)}{dt} & = & \omega_r {\alpha}^{(2)}_{g,1} (t) + g(t) {\alpha}^{(1)}_{e,0} (t) ,\nonumber \\
i \frac{d {\alpha}^{(2)}_{e,0} (t)}{dt} & = &\omega_0 {\alpha}^{(2)}_{e,0}(t) + g(t)  {\alpha}^{(1)}_{g,1} (t) ,\nonumber \\
i \frac{d {\alpha}^{(2)}_{e,1} (t)}{dt} & = &  \Omega_+ {\alpha}^{(2)}_{e,1}(t) +g(t)  {\alpha}^{(1)}_{g,0} (t).
\end{eqnarray}

\noindent
Substituting the value for the first order coefficients $\alpha^{(1)} (t) $, one can find the second order coefficients. The only non-zero coefficient is the following

\begin{eqnarray}
\label{coef1q2sol_pert}
{\alpha}^{(2)}_{g,0} \left( t \right)  &=& g_0^2 \frac{ \left\{  i \varpi_s \left[  2 i \Omega_+ t +\cos  \left( 2 \varpi_s t \right) -1  \right] -   \Omega_+ \sin  \left( 2 \varpi_s t \right)  \right\} }{ 4 \varpi_s \left[ \varpi_s +  \Omega_+ \right] \left[ \varpi_s -  \Omega_+ \right]  }  .
 \end{eqnarray}

\section{}
\label{appC}

The Laplace approach can simplify the solution of time-dependent differential equations. In this case, we consider the perturbative dynamics described by Eqs. (\ref{coeff1q_0_pert}), (\ref{coeff1q_1_pert}) and (\ref{coeff1q_2_pert}) and use the Laplace approach to find a solution. At zero-th order in $\delta$ we have

\begin{eqnarray}
\label{coeff1q_0_lap}
i  s {A}^{(0)}_{g,0} (s)  &= & 1 ,\nonumber \\ 
i s {A}^{(0)}_{g,1} (s) &= & \omega_r {A}^{(0)}_{g,1} (s) ,\nonumber \\
i s {A}^{(0)}_{e,0} (s)&= & \omega_0 {A}^{(0)}_{e,0} (s),\nonumber \\
i s {A}^{(0)}_{e,1} (s) &= &   \Omega_+ {A}^{(0)}_{e,1}(s)  ,
\end{eqnarray}

\noindent
where $A(s) = \mathcal{L}\left[ \alpha (t) \right](s)$ is the Laplace transform of the time-dependent coefficients of the wavefunction. Considering a system initially in the ground state $\lvert \psi (0) \rangle =   \lvert g, 0   \rangle $, the only non-zero coefficient is $ A^{(0)}_{g,0} (s) = 1/s$. Which gives $ \alpha^{(0)}_{g, 0} \left( 0 \right) = 1$.

\noindent
For simplicity we substitute the values for the zero-th order coefficients before taking the Laplace transform. At first order in terms of $\delta$, one finds 
 
\begin{eqnarray}
\label{coeff1q_1_lap}
i s {A}^{(1)}_{g,0} (s) &= & 0 ,\nonumber \\ 
i s {A}^{(1)}_{g,1} (s)&= & \omega_r {A}^{(1)}_{g,1}  (s)  ,\nonumber \\
i s {A}^{(1)}_{e,0} (s) &= & \omega_0 {A}^{(1)}_{e,0} (s)  ,\nonumber \\
i s {A}^{(1)}_{e,1} (s) &= &   \Omega_+ {A}^{(1)}_{e,1} (s) +G(s)  ,
\end{eqnarray}

\noindent
where $G(s) = g_0 \frac{s}{s^2 + \varpi_s^2}$ is the Laplace transform of the time-dependent coupling (\ref{g_sin}). The only non-zero coefficients at first order is

\begin{equation}
\label{coef1q1sol_lap}
 {A}^{(1)}_{e,1} \left( s \right) = - g_0  \frac{s}{s^2 + \varpi_s^2 } \frac{1}{is -  \Omega_+} ,
\end{equation}

\noindent
or in the time-domain

\begin{equation}
\label{coef1q1sol2_lap}
 {\alpha}^{(1)}_{e,1} \left( t \right) = - \frac{g_0}{2} \left[  \frac{e^{i \varpi_s t} }{\varpi_s -  \Omega_+ } - \frac{e^{i \varpi_s t} }{\varpi_s +  \Omega_+ }    - \frac{2 e^{- i  \Omega_+  t}  \Omega_+  }{\varpi_s^2 -  \Omega_+^2 } \right] .
\end{equation}

\noindent
At second order in terms of $\delta$, we have

\begin{eqnarray}
\label{coeff1q_2_lap}
i s {A}^{(2)}_{g,0} (s) &= & \mathcal{L}\left[ g(t)  {\alpha}^{(1)}_{e,1} \left( t \right) \right](s) ,\nonumber \\ 
i s {A}^{(2)}_{g,1} (s)&= & \omega_r {A}^{(2)}_{g,1}  (s)  ,\nonumber \\
i s {A}^{(2)}_{e,0} (s) &= & \omega_0 {A}^{(2)}_{e,0} (s)  ,\nonumber \\
i s {A}^{(2)}_{e,1} (s) &= &   \Omega_+ {A}^{(2)}_{e,1} (s) .
\end{eqnarray}

\noindent
The only non-zero coefficient is the following

\begin{equation}
\label{coef1q2sol_lap}
 {A}^{(2)}_{g,0} \left( s \right) = -\frac{i g_0^2 \left(s^4+ \left(4 \varpi_s^2+\Omega_+^2\right) s^2+3 i \varpi_s ^2 \Omega_+ s +2 \varpi_s ^2 \Omega_+^2\right)}{2 s \left(s^2+4 \varpi_s ^2\right) \Omega_+ (s-i (\varpi_s -\Omega_+)) (s+i (\varpi_s +\Omega_+))} ,
\end{equation}

\noindent
which in the time-domain gives

\begin{eqnarray}
\label{coef1q2sol2_lap}
{\alpha}^{(2)}_{g,0} \left( t \right)  &=& \frac{1}{16} g_0^2 \left(\frac{\varpi_s ^2 (6-4 i  \Omega_+ t)+2 i \Omega_+^2 (2 \Omega_+ t +7 i)}{(\varpi_s - \Omega_+)^2 (\varpi_s +\Omega_+)^2}+\frac{e^{-2 i  \varpi_s t }}{\varpi_s(\varpi_s -\Omega_+)}+\frac{e^{2 i  \varpi_s t }}{\varpi_s(\varpi_s +\Omega_+)}\right) .
 \end{eqnarray}


\begin{thebibliography}{99} 

\bibitem{rahav} S.Rahav, I. Gilary, and S. Fishman, Phys. Rev. A {\bf 68}, 013820 (2003).

\bibitem{shtoff} A. V. Shtoff and Yu. Yu. Dmitriev, Optics and Spectroscopy {\bf 102}, 166 (2007).

\bibitem{goldman} N. Goldman and J. Dalibard, Phys Rev X {\bf 4}, 031027 (2014).

\bibitem{holthaus} M. Holthaus, Journal of Physics B {\bf 49}, 013001 (2015).

\bibitem{chu} S. I. Chu and D. A. Telnov, Physics reports {\bf 390}, 1-2 (2004).

\bibitem{floquet} G. Floquet, Ann. de l'Ecole Norm. Sup. {\bf12}, 47 (1883).

\bibitem{eastham} M. S. P. Eastham, \textit{The spectral theory of periodic differential equations} (Scottish Acad. Press, 1973).

\bibitem{daleckii} Ju. L. Daleckii and M. G. Krein, \textit{Stability of solutions of differential equations in Banach space}, Transl. Math. Monogr., 43, Amer. Math. Soc. (1974).

\bibitem{ashcroft} N.W. Ashcroft, N.D. Mermin, \textit{Solid State Physics} (Holt, Rinehart and Winston, 1976).

\bibitem{shirley}  J. H. Shirley, Phys. Rev. {\bf138}, 979 (1965).

\bibitem{sambe}  H. Sambe, Phys. Rev. A  {\bf 7}, 2203 (1973).

\bibitem{levante} T. O. Levante, M. Baldus, B. H. Meier, and R. R. Ernst, Molecular Physics \textbf{86}, 5 (1995).

\bibitem{leskes} M. Leskes, P. K. Madhu, and S. Vega, Progress in nuclear magnetic resonance spectroscopy \textbf{57}, 4 (2010).

\bibitem{grifoni} M. Grifoni and T. H\"{a}nggi, Phys. Rep. {\bf304}, 229 (1998).

\bibitem{amico2} M. Amico, O. L. Berman, and R. Ya. Kezerashvili,  \pra {\bf 98}, 042325 (2018).

\bibitem{amico3} M. Amico, O. L. Berman, and R. Ya. Kezerashvili, Phys. Lett. A  {\bf 383}, 487-493 (2019).

\bibitem{shapiro} D. S. Shapiro, A. A. Zhukov, W. V. Pogosov, and Yu. E. Lozovik, Phys. Rev. A \textbf{91}, 063814 (2015).

\bibitem{zhukov}  A. A. Zhukov, D. S. Shapiro, W. V. Pogosov, and Yu. E. Lozovik, Phys. Rev. A \textbf{93}, 063845 (2016).

\bibitem{amico1} M. Amico, O. L. Berman, and R. Ya. Kezerashvili, \pra {\bf 96}, 032328 (2017).

\bibitem{remizov} S. V. Remizov, A. A. Zhukov, D. S. Shapiro, W. V. Pogosov, and Yu. E. Lozovik, Phys. Rev. A \textbf{96}, 043870 (2017).

\bibitem{amico4} M. Amico, O. L. Berman, and R. Ya. Kezerashvili,   \pra {\bf 100}, 013841 (2019).

\bibitem{dle} N. B. Narozhny, A. M. Fedotov, and Yu. E. Lozovik, Phys. Rev. A \textbf{64}, 053807 (2001).

\bibitem{son} S. K. Son, S. Han, and S. I. Chu, \pra {\bf79}, 3 (2009).

\bibitem{deng} C. Deng, J. L. Orgiazzi, F. Shen, S. Ashhab, and A. Lupascu,  \prl {\bf 115}, 13 (2015).

\bibitem{pirkka} J. M. Pirkkalainen, S. U. Cho, J. Li, G. S. Paraoanu, P. J. Hakonen, and M. A. Sillanp\"{a}\"{a}, Nature {\bf 494}, 7436 (2013).

\bibitem{marinescu}  M. Marinescu and M. Gavrila, Phys. Rev. A {\bf 53}, 2513 (1996).

\bibitem{gavrila} M. Gavrila, \textit{Atoms in Intense Laser Fields} (Academic Press, New York, 1992), pp. 435-510.

\bibitem{creffield} C. E. Creffield and G. Platero, Phys. Rev. B {\bf65}, 113304 (2002).

\bibitem{qutip1} J. R. Johansson, P. D. Nation, and F. Nori, Comput. Phys. Commun. {\bf 183}, 1760 (2012).

\bibitem{sakurai} J. J. Sakurai, \textit{Modern Quantum Mechanics, Revised Edition} (Addison-Wesley, Reading, MA,1994).

\bibitem{trotter} H. F. Trotter, Proc. Am. Math. Soc. {\bf10}, 545-551 (1959).

\bibitem{suzuki1} M. Suzuki, Commun. Math. Phys. {\bf51}, 183-190 (1976).

\bibitem{suzuki2} M. Suzuki, Progr. Theor. Phys. {\bf56}, 1454-1469 (1976).

\bibitem{poulin} D. Poulin, A. Qarry, R. Somma, and F. Verstraete, \prl {\bf 106}, 17 (2011).

\bibitem{lamata} L. Lamata, Scientific Reports {\bf 7}, 43768 (2017).

\bibitem{qutip2} J. R. Johansson, P. D. Nation, and F. Nori, Comput. Phys. Comm. {\bf184}, 1234 (2013).





%
%
%
%
%
%
%
%
%
%
%
%
%
%
%
%
%
%
%
%
%
%
%
%
%

\end{thebibliography}
\end{document}